\documentclass[conference]{IEEEtran}
\IEEEoverridecommandlockouts
\usepackage{amsmath,amssymb,amsfonts}
\usepackage{algorithmic}
\usepackage{caption}
\usepackage{subcaption}
\usepackage{graphicx}
\usepackage{textcomp}
\usepackage{gensymb}
\usepackage{xcolor}
\usepackage{siunitx}
\usepackage{glossaries}
\usepackage{url}
\newacronym{med}{MED}{mean Euclidean distance}
\newacronym{awgn}{AWGN}{additive white gaussian noise}
\newacronym{snr}{SNR}{signal-to-noise ratio}
\newacronym{sir}{SIR}{signal-to-interference ratio}
\newacronym{agc}{AGC}{automatic gain control}
\newacronym{cdf}{CDF}{cumulative distribution function}
\newacronym{ml}{ML}{machine learning}
\newacronym{dl}{DL}{deep learning}
\newacronym{ism}{ISM}{image-source model}
\newacronym{rir}{RIR}{room impulse response}
\newacronym{slam}{SLAM}{simultaneous localization and mapping}
\newacronym{en}{EN}{energy neutral}
\newacronym{cnn}{CNN}{convolutional neural network}
\newacronym{tof}{ToF}{time of flight}
\newacronym{ips}{IPS}{indoor positioning system}
\usepackage[english]{babel}
\usepackage[backend=biber,style=ieee,dashed=false]{biblatex }
\AtBeginBibliography{\footnotesize}
\usepackage{tabularx}
\newcolumntype{Y}{>{\centering\arraybackslash}X} 
\newcolumntype{L}{>{\raggedright\arraybackslash}X}
\newcolumntype{R}{>{\raggedleft\arraybackslash}X}
\usepackage{cellspace}
\usepackage{multicol}
\usepackage{multirow}
\usepackage{tablefootnote}
\usepackage{pifont}
\newcommand{\xmark}{\ding{55}}%

\usepackage{url}

\usepackage[super]{nth}

\begin{document}

\title{An Acoustic Simulation Framework to Support Indoor Positioning and Data Driven Signal Processing Assessments\\
}

\author{\IEEEauthorblockN{Daan Delabie, Chesney Buyle, Bert Cox, Liesbet Van der Perre, Lieven De Strycker}
\IEEEauthorblockA{KU Leuven, WaveCore, Department of Electrical Engineering (ESAT), KU Leuven Ghent, 9000 Ghent, Belgium \\
daan.delabie@kuleuven.be}}

\maketitle

\begin{abstract}
We present an indoor acoustic simulation framework that supports both ultrasonic and audible signaling. The framework opens the opportunity for fast indoor acoustic data generation and positioning development. The improved Pyroomacoustics-based physical model includes both an \gls{ism} and ray tracing method to simulate acoustic signaling in geometric spaces that extend typical shoe-box rooms. Moreover, it offers the convenience to facilitate multiple speakers and microphones with different directivity patterns. In addition to temperature and air absorption, the room reverberation is taken into account characterized by the RT60 value or the combination of building materials. Additional noise sources can be added by means of post processing and/or extra speakers. Indoor positioning methods assessed in simulation are compared with real measurements in a testbed, called `Techtile'. This analysis confirms that the simulation results are close to the measurements and form a realistic representation of the reality. The simulation framework is constructed in a modular way, and parts can be replaced or modified to support different application domains. The code is made available open source.
\end{abstract}


\section{Introduction}
Acoustic signaling can serve a very broad set of applications ranging from localization to sonar, speech recognition, and sound classification to imaging in, for example, the medical world. In addition to audible acoustic signals, ultrasound is gaining attention, more specifically in the domain of positioning. It possesses a low propagation speed compared to RF and light, and lies outside the audible frequency range of the human ear. Moreover, ultrasound and acoustic waves are suitable to provide time-based distance estimations for indoor positioning with low-cost hardware.
Previous research presented a hybrid RF-acoustic \gls{ips} using ultrasonic chirp signals for ranging estimations between speaker anchors and a mobile microphone~\cite{mdpi_bert}. In order to achieve higher accuracy and reliability at all locations in an indoor environment, new positioning algorithms need to be developed and tested. \Gls{ml} models offer the opportunity to provide significant improvements~\cite{surveyML, survey_ML_acoustics}. The requirement for large data sets to train such models is a major hurdle: acoustic data collection for indoor positioning or audio classification is a time consuming task. Acoustic signal propagation simulators in combination with real-life measurements can provide a solution. For this reason, we developed a simulation framework that supports acoustic and ultrasonic signals combined with an option to extend it with extra building blocks for \gls{ml}, positioning or classification algorithms. Hence, the flow from data creation to analysis of the operation of specific applications is fully supported within one framework, which fosters smooth development.


Acoustic simulation packages are widely available since the early 1960s~\cite{book_acoustics}. Popular commercial examples include: Odeon Room Acoustics Software, CATT, EASE, Siemens Acoustic Simulation and OpenFOAM. Most are user-friendly though expensive, incapable to apply ultrasonic signals in the simulated rooms or cannot be easily expanded with desired post-processing~\cite{overview_new}. There are also open-source toolboxes and libraries such as `Roomsim'~\cite{roomSimMatlab} for Matlab and `Pyroomacoustics'~\cite{pyroomacoustics} for Python. The latter is used as the basis for the proposed indoor acoustic simulation framework. A faster and more accurate shoe-box room simulator than `Roomsim' was proposed in~\cite{a_fast_and}. This simulator takes into account air absorption, humidity, temperature, scattering, and directivity. Nevertheless, only a shoe-box model simulation is possible. \Gls{cnn} based sound source localization is performed in~\cite{CNNsoundsource}. Pyroomacoustics is used for data creation to feed the \gls{cnn}. As with the other proposed simulation frameworks, no complete extension is provided for ultrasonic signals. 

One of the few simulators that supports ultrasonic signals is Field II. Closely related to the research presented in this paper, an ultrasonic chirp was emitted and received, followed by a cross-correlation for range estimation within a small room~\cite{field2_1}. Unfortunately, the reflection/reverberation property of the room, which noticeably affects the accuracy and reliability of an \gls{ips}, is not taken into account~\cite{field2_2}.


Since a comprehensive and representative acoustic simulator for both audible audio and ultrasound did not yet exist to the best of our knowledge, we adapted the open-source Pyroomacoustics library. In addition to the audible sound, ultrasonic signals in the range of 20 to 50 kHz can be modelled. This additional range matches the frequency response of commercially available ultrasonic (MEMS) microphones. A chirp-based \gls{ips}, closely related to~\cite{mdpi_bert}, is simulated in our presented framework as a means for validation. Data generation, post-processing, positioning and evaluation take place within the same framework. This framework can readily be used to test and compare acoustic/ultrasonic indoor positioning algorithms where both accuracy and reliability can be checked. Nevertheless, with some adaptations, the framework can also be used to generate audio data sets for other applications, e.g., speech recognition models or sound classification in general. The post-processing can be altered to output the audio fragment and the positioning engine can be replaced by, e.g., a classification engine. In addition, jammers and noise can be added. 

The paper is structured as follows. An overview of the proposed simulation framework is given in Section~\ref{sec:sim_framework}. This reflects what is currently implemented, with focus on the analysis of indoor positioning algorithms supporting compatibility with previous research~\cite{mdpi_bert}. Afterwards, a real world validation is presented in Section~\ref{sec:comp}, followed by a discussion with referenced future work in Section~\ref{sec:discussion}. Lastly, a conclusion is presented in Section~\ref{sec:conclusion}.

\section{Simulation Framework}\label{sec:sim_framework}
\begin{figure}[!htb]
\centering
    \centering
    \includegraphics[width=0.45\linewidth]{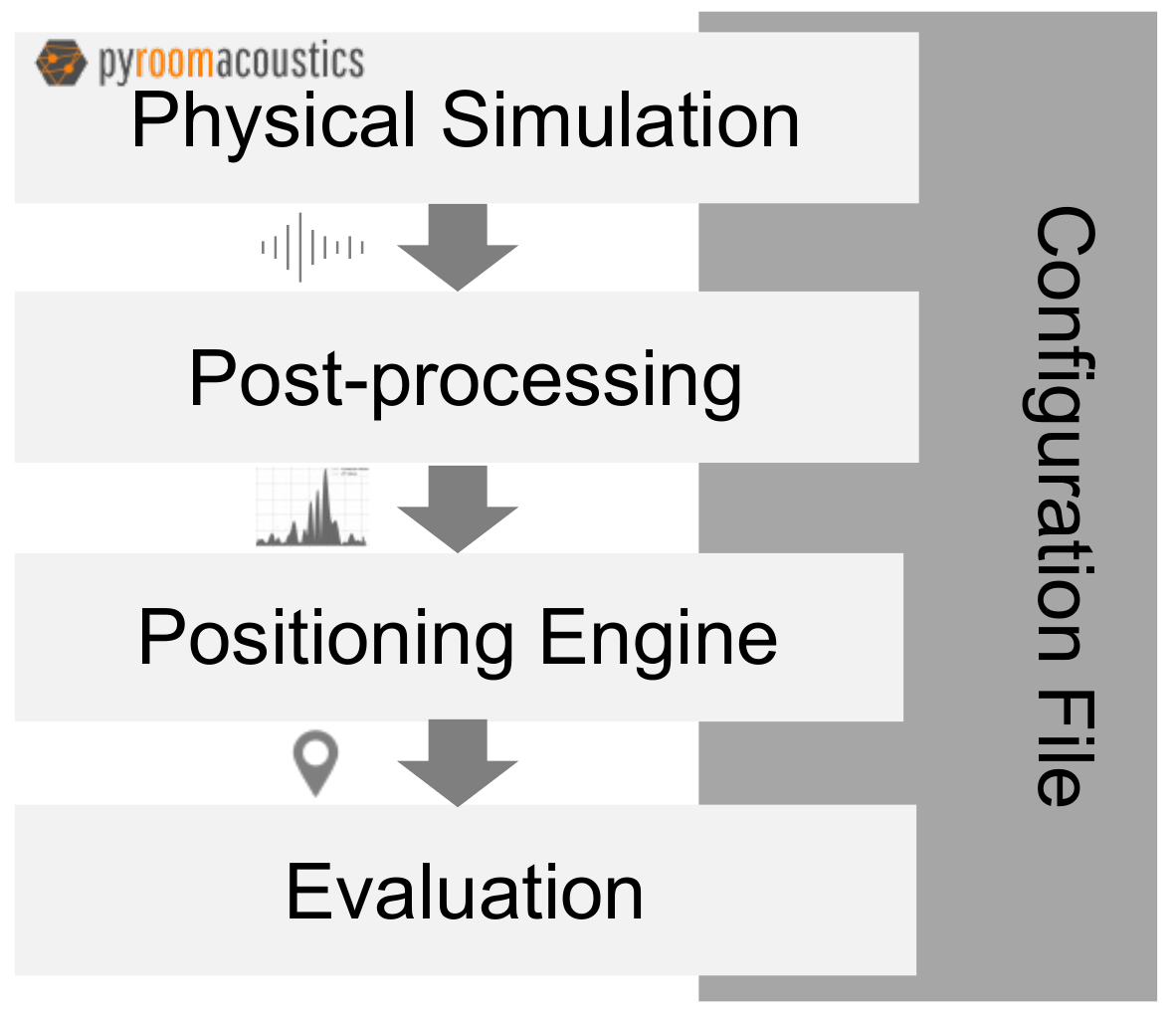}%
    \caption{Overview of the acoustic positioning simulation framework.}%
    \label{fig:overview_framework}
\end{figure}

The proposed simulation framework\footnote{\url{github.com/DaanDelabie/AcousticSimulationFramework}} consists of four main parts, as presented in Fig.~\ref{fig:overview_framework}: the physical simulator, post-processing layer, the positioning engine and a final evaluation layer. 

\setlength{\belowcaptionskip}{-7pt}
\setlength{\belowcaptionskip}{0pt}

\begin{figure}[!t]
\centering
    \centering
    \includegraphics[width=0.9\linewidth]{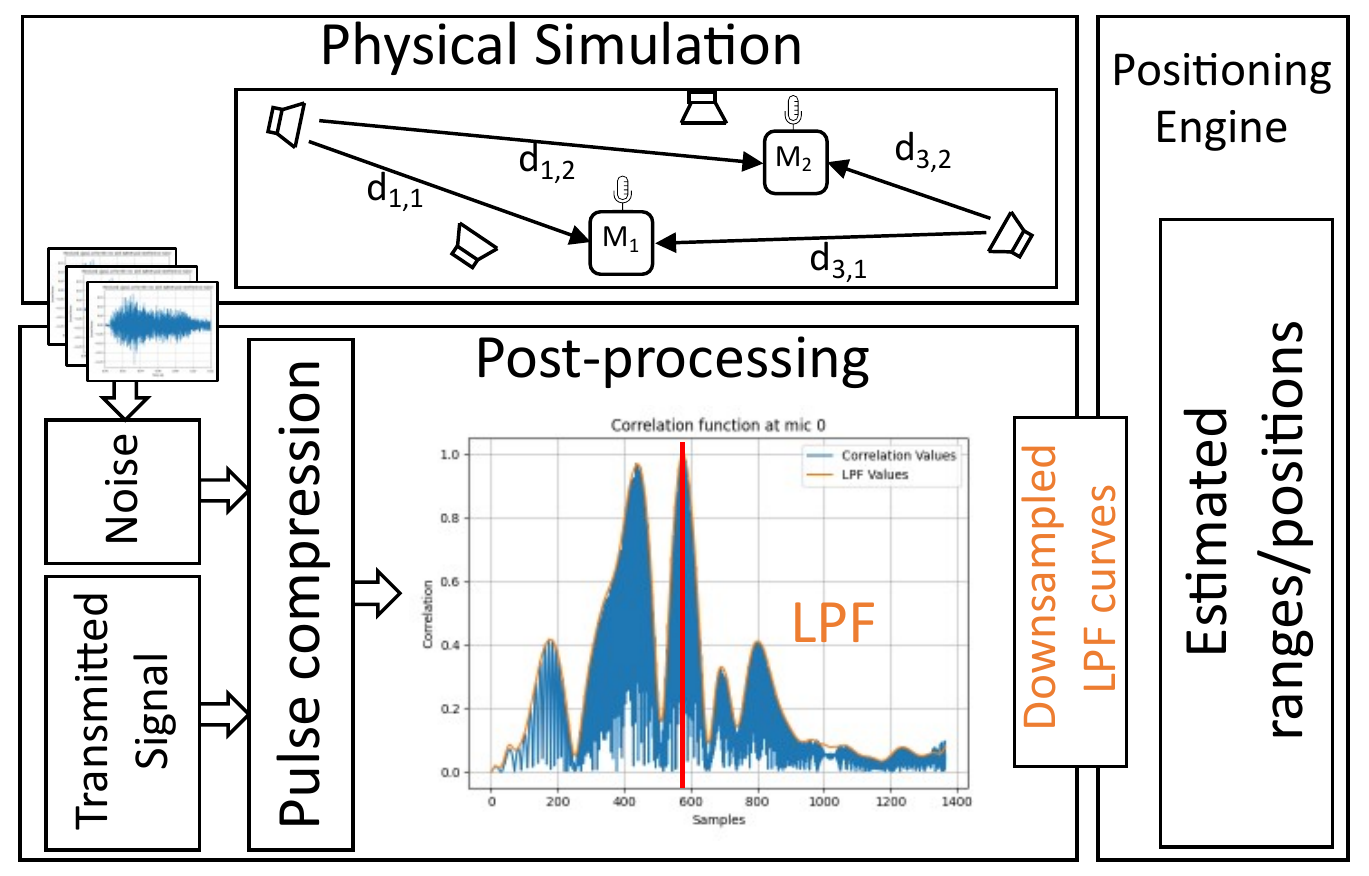}%
    \caption{Simulation flow for the analysis of the \gls{ips}.}%
    \label{fig:sim_flow}
\end{figure}

The top layer of this approach (i) models an indoor space, (ii) processes the physical properties of the space and (iii) calculates the propagation of acoustic signals. The addition of noise and processing of the received signals, e.g., pulse compression, is performed in the post-processing layer. This layer feeds the positioning engine where based on the input the 3D positioning of objects or sources is performed. To compare different algorithms in all three previous layers, a separated evaluation section has been added. All these components are linked through common parameters, bundled in a single configuration file which defines the desired simulation. The fact that the simulation framework is composed of stand-alone building blocks makes adaptations and comparisons on a component-by-component basis straightforward and efficient. For example, when comparing different positioning algorithms, the output of the physical simulation part and the post-processing part can be reused without recalculation, resulting in a great profit of time and computing resources. If a comparison needs to be made across different \gls{snr} values, only the simulation part from post-processing onward should be retaken. Moreover, the content and thus inputs and outputs of these blocks can easily be adapted to support other applications or algorithms. To describe what is possible with this simulator, each building block is discussed in more detail.  An ultrasonic \gls{ips}~\cite{mdpi_bert}, with speakers that act as anchor nodes and microphones as mobile devices, is used as a tutorial. The principle of the implemented analysis method is illustrated in~Fig~\ref{fig:sim_flow}.

\setlength{\belowcaptionskip}{-10pt}
\setlength{\belowcaptionskip}{0pt}

\subsection{Physical Simulation}
This building block is based on the Pyroomacoustics library~\cite{pyroomacoustics}, with minor adjustments to support multi-processing and ultrasonic signals. The properties of this component are summarised.

\begin{itemize}
    \item Multiple rooms can be simulated in parallel on different processor cores, allowing parallel/multi-processed testing of multiple speaker positions that are not allowed to output simultaneously.
    \item  The physical simulation extends conventional shoe-box models and can create any desired flat ceiling 3D room shape. This simulation environment is based on the given ground plane vertices, heights and room materials for the accompanying ceiling, floor and walls.
    \item The option to define an RT60 value, the time spent for the \gls{rir} to decay by 60 dB, instead of using materials is also offered. Unlike using materials, the RT60 value can only be used when simulating a shoe-box configuration since it uses the inverse Sabine formula~\cite{sabine} to approximate the wall energy absorption.
    \item Microphones and speakers can be placed freely at designated locations. In the example setup, as mentioned in the introduction and illustrated in Fig~\ref{fig:sim_flow}, the indoor positioning of a microphone is determined based on chirp signals transmitted at anchor nodes (speakers) located on the walls and ceiling. To assess the localization performance throughout the room, an equally spaced grid of microphone locations is generated. To train \gls{ml} models, a split by a certain percentage of the training, development and test set is typically required. The train-dev-test set ratios are included in the configuration file, which is automatically taken into account. In case \gls{ml} techniques are applied for the indoor positioning algorithms, the equally spaced location set is used as a test set, so that well-organized figures are created and can be compared. For the train and development set, a random set of locations is generated that is located in between the test set grid. These concepts are shown in Fig.~\ref{fig:position_generation}.

    \setlength{\belowcaptionskip}{-10pt}
    \begin{figure}[t]
    \centering
    \centering
    \includegraphics[width=0.9\linewidth]{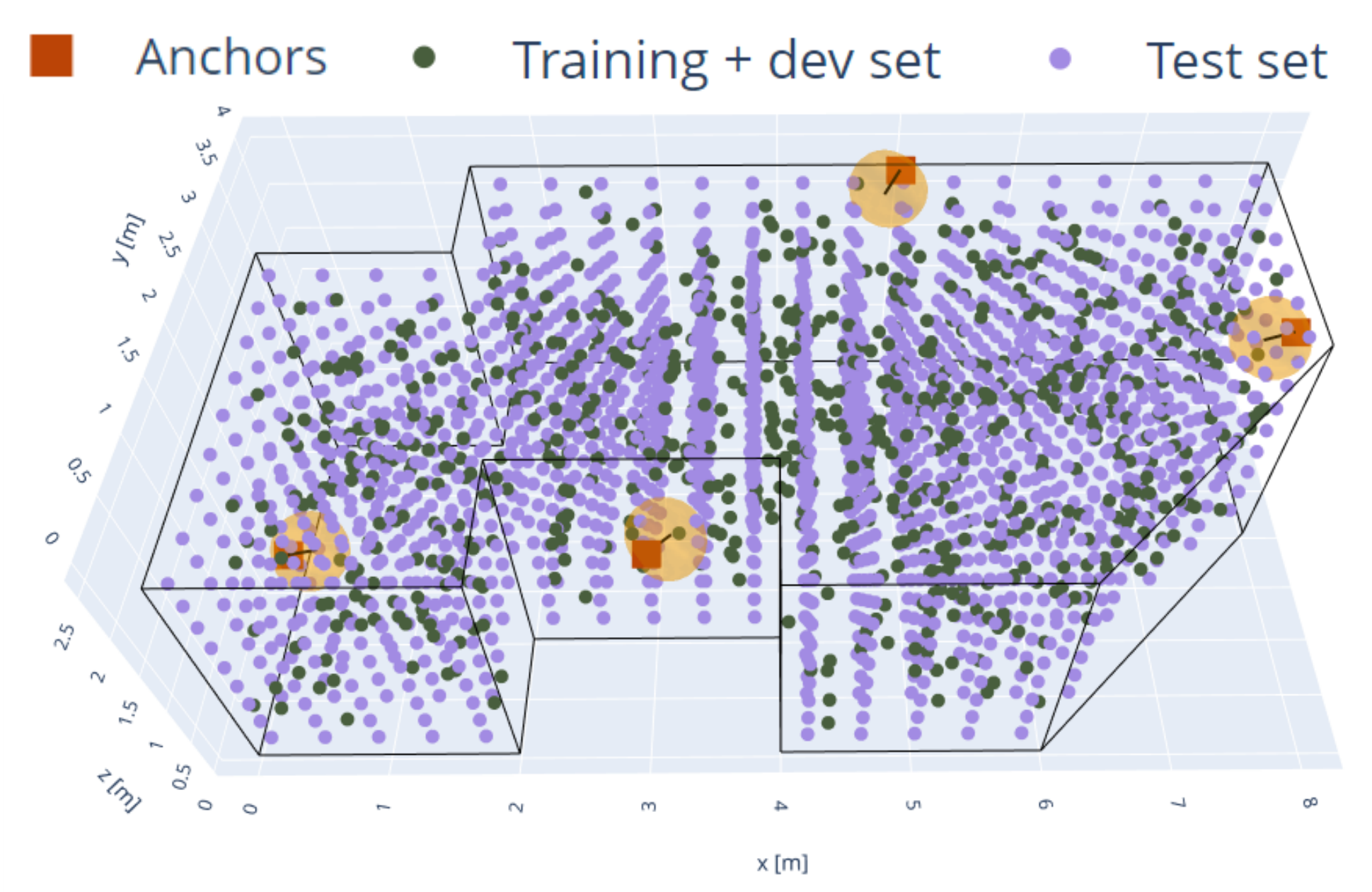}%
    \caption{An example of generated test grid and train/dev cloud microphone positions in a random shaped room with speakers representing anchors.}%
    \label{fig:position_generation}
    \end{figure}
    \setlength{\belowcaptionskip}{0pt}

    \item The directivities and the transmit signal can be set as desired. \item The speed of sound in air is determined by the temperature $\theta$ specified ~in~$\mathrm{^{\circ} C}$:

    \begin{equation}
        v_{sound} = 331.3 \, \sqrt{\frac{273.15 + \theta}{273.15}} = 20 \, \sqrt{273.15 + \theta}
    \end{equation}

    \item To perform the actual propagation simulation, an \gls{ism} and ray tracing option are provided by Pyroomacoustics, each with their (current) capabilities as summarized in Table~\ref{table:ism-ray}.

    \setlength{\belowcaptionskip}{-5pt}
    {\renewcommand{\arraystretch}{1.5} 
\begin{table}[h]
\centering
    \begin{tabularx}{\linewidth}{>{\hsize=1\hsize}L >{\hsize=1\hsize}c >{\hsize=1\hsize}c>{\hsize=1\hsize}c>{\hsize=1\hsize}c>{\hsize=1\hsize}c>{\hsize=1\hsize}c>{\hsize=1\hsize}c>{\hsize=1\hsize}c}
        \hline
        
        \multicolumn{1}{l}{} & \multicolumn{4}{c}{\textbf{Shoebox}} & \multicolumn{4}{|c}{\textbf{Non-Shoebox}} \\
        
        \multicolumn{1}{l}{} & \multicolumn{2}{c}{\textbf{Materials}} & \multicolumn{2}{c}{\textbf{Directivities}} & \multicolumn{2}{|c}{\textbf{Materials}} & \multicolumn{2}{c}{\textbf{Directivities}} \\
        \hline         
        \multicolumn{1}{l}{} & \multicolumn{1}{c}{\textbf{$\ne$}} & \multicolumn{1}{c}{\textbf{$=$}} & \multicolumn{1}{|c}{\textbf{Mic}} & \multicolumn{1}{c}{\textbf{Speaker}} & \multicolumn{1}{|c}{\textbf{$\ne$}} & \multicolumn{1}{c}{\textbf{$=$}} & \multicolumn{1}{|c}{\textbf{Mic}} & \multicolumn{1}{c}{\textbf{Speaker}}\\
        \hline     
          ISM & \checkmark & \checkmark & \checkmark & \checkmark & \multicolumn{1}{|c}{\xmark} & \checkmark & \checkmark & \xmark\\
          RT &  \xmark & \checkmark & \xmark & \xmark & \multicolumn{1}{|c}{\xmark} & \checkmark & \xmark & \xmark\\                          
        \hline
    \end{tabularx}
    \caption{Possibilities given an \gls{ism} or raytracing (RT) configuration applied to a (non)-shoebox shaped room. Materials on the walls, ceiling and floor can be equal ($=$) or different ($\ne$). The supported cardioid family directivities are frequency-independent.}
	\label{table:ism-ray}
\end{table}
}

    \setlength{\belowcaptionskip}{0pt}
    
\end{itemize}

While the Pyroomacoustics package has been promoted as a tool for signal processing in the audio domain, the simulation environment can be adapted for ultrasonic signal propagation as well. However, two parameters should be adjusted to emulate the propagation of ultrasonic signals in real indoor environments: air absorption and material absorption.

\begin{itemize}
    \item As acoustic signals propagate through air, the waves get attenuated due to viscosity and relaxation processes~\cite{pierce2019acoustics}. This results in an exponential decrease in the intensity of the acoustic signal. In Pyroomacoustics and thus the adapted simulator, air absorption is modelled as a factor $\alpha_{\mathrm{air\,decay}}$:

    \begin{equation}
        \alpha_{\mathrm{air\,decay}} = e^{-\alpha_{\mathrm{abs}} d}, 
    \end{equation}

    where $\alpha_{\mathrm{abs}}$ is the acoustic absorption coefficient in Np/m and $d$ is the distance in m. The acoustic absorption coefficient $\alpha_{\mathrm{abs}}$ depends to a great extend on the frequency of the acoustic signal, and on environmental parameters such as temperature and relative humidity. Pyroomacoustics supports only frequency-dependent absorption coefficients values for octave bands between $\SI{125}{\hertz}$ and $\SI{8}{\kilo\hertz}$. Specific values for the ultrasonic frequency range are manually added to the newly adapted simulator, according to previous research performed in~\cite{rekhi2017wireless, bass1995atmospheric}. 

    \item Acoustic waves are also attenuated when reflected off objects and walls. The amount of energy that is absorbed depends, among other things, on the materials. The proposed simulator models materials by means of an energy absorption coefficient $\alpha_{\mathrm{abs,material}}$, which represents the ratio of the energy absorbed by the material to the total energy of the incident sound wave. Consequently, the amplitude of the reflected wave $A_{\mathrm{reflected}}$ can be calculated by

    \begin{equation}
        A_{\mathrm{reflected}} = A_{\mathrm{incident}} \sqrt{1 - \alpha_{\mathrm{abs,material}}},    
    \end{equation}
    
    where $A_{\mathrm{incident}}$ is the amplitude of the incident acoustic wave and $\alpha_{\mathrm{abs,reflection}}$ is the energy absorption coefficient of a specific material. A material can also be characterized by multiple energy absorption coefficients to model its frequency-dependent behaviour. As in the case of air absorption, frequency-dependent wall absorption is originally supported in Pyroomacoustics, but only for the octave bands between $\SI{125}{\hertz}$ and $\SI{8}{\kilo\hertz}$. Absorption coefficients are added in the simulator extension to support ultrasonic signals. However, to the best of our knowledge, very little research is available on absorption for several materials in the ultrasound frequency range. The usually employed impedance tube to determine the absorption coefficient can unfortunately not be used in the case of ultrasonic signaling, given the necessary strict and physically impossible-to-use dimensioning. To obtain an indication of the ultrasonic absorption coefficient of plywood, we measured the intensities of the direct and reflected wave within a frequency range of 20 to 50 kHz. The used method is based on the ISO 13472-1 standard. First, a free-field calibration is performed between a speaker and microphone to check the intensity of the incident sound at a certain distance for a certain frequency. Afterwards, a wooden panel is placed at the same distance to receive the reflected short pulse with a certain frequency at the position of the speaker. Since unwanted reflections are avoided when sending the short pulse, the reflected pulse can be purely received within a different time window than the one of the transmitted pulse. Within the specified frequency domain, an average absorption coefficient of 0.43 was obtained with a standard deviation interval of $\mathrm{[0.14, 0.72]}$.
    
\end{itemize}


The physical simulation part of the framework outputs the corresponding \glspl{rir} and received audio samples. Afterwards, the audio samples are downsampled to meet the desired sampling frequency of the microphone. This data is passed on to the post-processing part.

\subsection{Post-Processing}
The post-processing part can add different types of noise and interference to the signal depending on a pre-specified \gls{snr} and \gls{sir} values. This allows to determine the influence of noise on, for example, the accuracy of the \gls{ips} under test. Interference can also be obtained by adding an extra speaker in the physical model, making it inherently present in the audio signal. The disadvantage of this method is that a modified noise parameter impacts the simulation time since this is performed in the physical part.

This post-processing block is easily extendable with necessary additions such as a Monte-Carlo simulation to assess the influence of noise at different SNR levels. In addition, downsampling, \gls{agc}, pulse compression and filtering can be used. To give an example of what is possible, the previously illustrated \gls{ips} as shown in Fig~\ref{fig:sim_flow} is elaborated, to comply with the procedure of~\cite{mdpi_bert}. After receiving the signals from the physical simulation, noise is added, followed by pulse compression based on a transmitted chirp signal and the noisy received part of that chirp signal. Afterwards, a low-pass filter is added to obtain the envelope curve of the pulse compression. The index of the peak of this curve gives an indication of the range between the corresponding speaker and microphone. This envelope curve is downsampled again to create a fixed output size among different simulation setups, for example, to use this as input features of a \gls{cnn}.

\subsection{Positioning Engine}
The positioning engine is used to perform a location estimate based on the given input data. In the exemplary configuration, using the peak prominence algorithm applied to the envelope of the pulse compression, a \gls{tof} and thus ranging estimation is performed between a certain speaker and microphone. The range estimates are then used to estimate a 3D position, for example using a least squares method. As mentioned in the physical simulation section, there is a possibility to already provide a test, train and dev data split. Since \gls{ml} plays an increasingly important role in indoor positioning, it releases the possibility to apply and learn different models to obtain better accuracy and reliability of the 3D position estimate. This simulation model was developed for future research to investigate \gls{ml} techniques for ultrasonic signal-based indoor positioning.

\subsection{Evaluation}
\begin{figure}[tb!]
\centering
    \centering
    \includegraphics[width=0.9\linewidth]{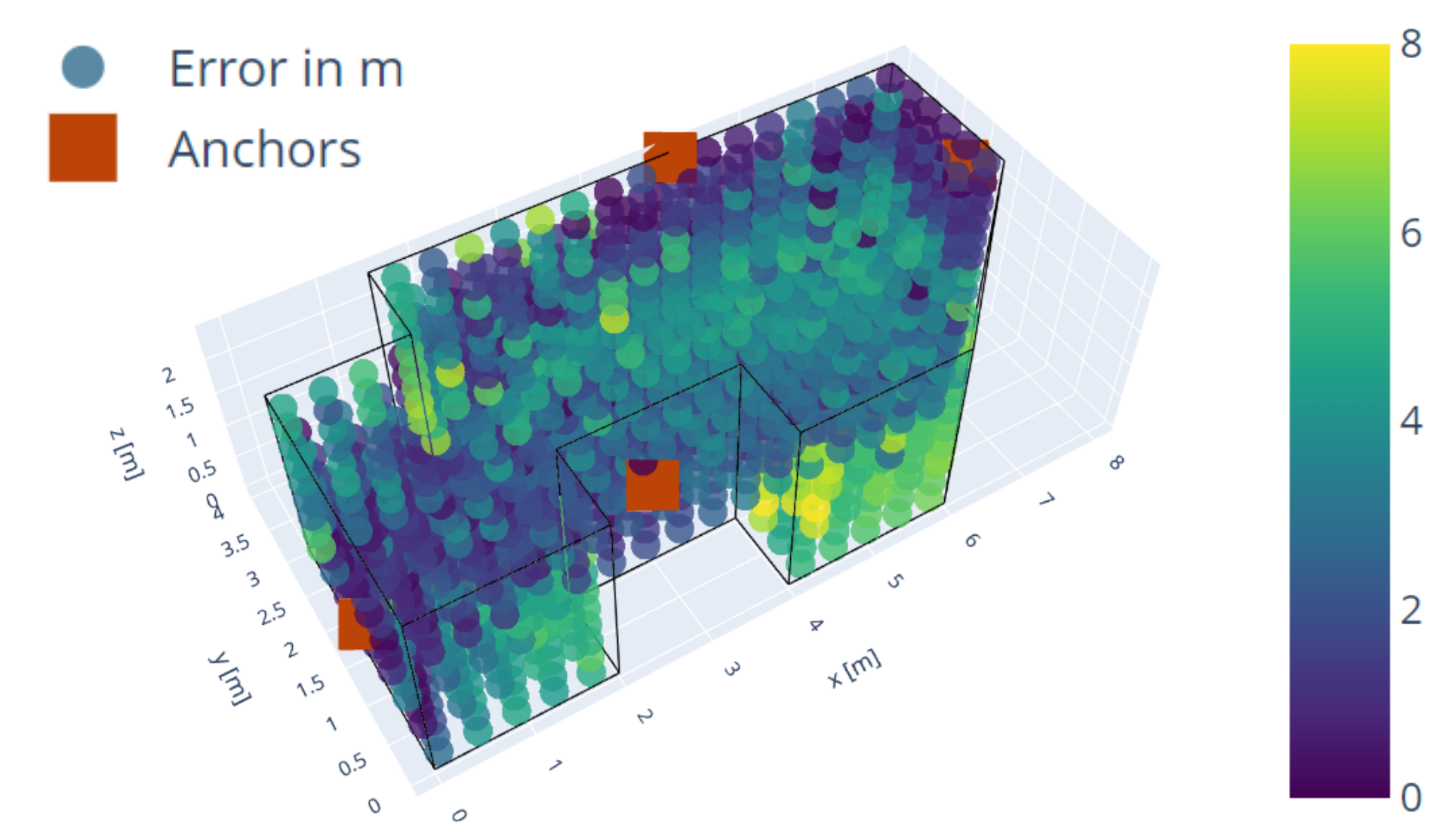}%
    \caption{An example of the Euclidean distance error on different locations in a 3D room.}%
    \label{fig:3dplot_ex}
\end{figure}

A graphical representation of the results can lead to better interpretation and can be important when designing applications. If the simulation framework is used for indoor positioning, the current evaluation content can be used to report \glspl{cdf} and automated 3D plotting independent of the shape of the room to display the accuracy of the position estimate as a function of the location in space. An example is given in Fig.~\ref{fig:3dplot_ex}. This part is very application dependent and can, as a result, be easily adapted to visualise the answers to the specific research questions.

\setlength{\belowcaptionskip}{-10pt}
\setlength{\belowcaptionskip}{0pt}

\section{Real world validation} \label{sec:comp}
To validate the proposed simulation model for its physical world representation capabilities, a comparison between a simulation and a measurement set is performed. The measured sets are gathered in a highly reverberant and thus acoustically challenging environment~\cite{IPINBert}. Since the main focus of this framework lies on acoustic indoor positioning, the same positioning algorithms are applied to the simulated and measured data set and the corresponding \glspl{cdf} are compared. The simulation framework was configured to emulate the real world measurement environment~\cite{IPINBert} as good as possible.

Both the real-life and simulation setup consist of four speaker anchors targeting a microphone which is placed at 150 different positions inside the `Techtile' testbed~\cite{techtile-acoustic}. The speakers are located at positions [0.5, 0.05, 0.146]m, [1.015, 3.950, 2.215]\,m, [5.183, 0.067, 0.744]\,m and [6.901, 3.948, 1.322]\,m. Positioning signals, consisting of a linear, descending chirp of 30\,ms, between 45\,KHz-25\,KHz, are emitted sequentially by the anchor nodes. The testbed is simulated as a wooden room with two open sides, emulated as absorbing surfaces within an \gls{ism}. The absorption co\"{e}fficient of wood is set to 0.45 in the adapted Pyroomacoustics library since ultrasound is addressed. Given the aforementioned measurements on a wooden Techtile panel within this measurement's frequency range, this value can be considered. The air absorption for ultrasound is also taken into account.

\begin{figure}[t!]
\centering
    \centering
    \includegraphics[width=0.9\linewidth]{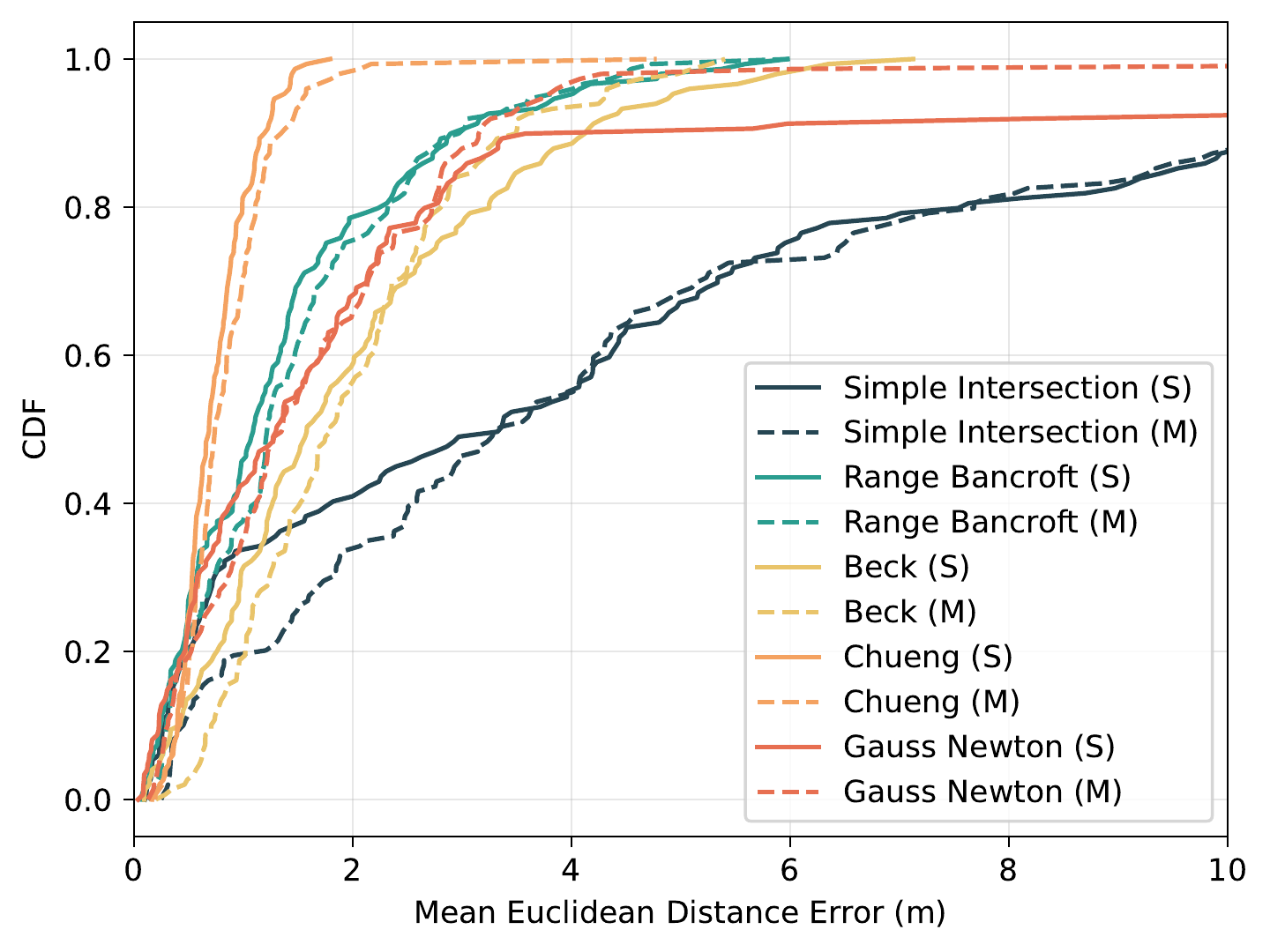}%
    \caption{Comparison between the measured (M) and simulated (S) CDFs of the \gls{med} error for different position estimation algorithms.}%
    \label{fig:comparison_figure}
\end{figure}

 The microphone samples the received signal during 1~ms at a sampling frequency of 250~kHz as described in~\cite{IPINBert}. The directivities of both speaker and microphone are also taken into account. An omni-directional pattern and a hypercardioid pattern are chosen for the microphone and speakers respectively. The speakers point towards the center of the room. The SNR value is assumed to be 30~dB. For range estimation, the maximum value of the bit based pulse compression is used, as described in~\cite{IPINBert}. After this ranging estimate between the microphone at a certain position and 4 speakers/anchors, a location estimate is determined, in this case, using the following acclaimed algorithms: simple intersections, range Bancroft, Beck, Chueng and Gauss-Newton. Fig~\ref{fig:comparison_figure} shows the comparison between simulated and measured values of the \glspl{cdf} obtained by the positioning algorithms. This demonstrates that the results achieved with the simulation framework closely match the reality for this indoor positioning application.

\setlength{\belowcaptionskip}{-10pt}
\setlength{\belowcaptionskip}{0pt}

\section{Discussion and future work}\label{sec:discussion}
The simulation environment makes it possible to quickly test \glspl{ips} and generate acoustic datasets. This framework will mainly be used to increase the accuracy and reliability of \glspl{ips} within the entire (not necessarily shoe-box) room. To this end, distributed anchor deployments and \gls{ml} techniques are being studied to extend the already existing hybrid-acoustic positioning system for \gls{en} devices~\cite{mdpi_bert}. If the simulation has to provide a lot of sequential speaker outputs at different locations, the simulation time can increase considerably, even in a multi-processed manner. Multiple microphones can receive sound within the same simulation episode, so that for example the simulation performed in Section~\ref{sec:comp} only took about 10 minutes on an Intel 8th Gen octa-core i7 processor. A disadvantage of this simulation framework is that no objects can be placed in the room at the moment. On the other hand, this framework offers flexibility and detailed possibilities to mimic a real environment.

\section{Conclusion} \label{sec:conclusion}
An \gls{ism} and ray tracing simulation framework, based on the Pyroomacoustics library, was proposed making it possible to generate acoustic/ultrasonic indoor data sets and to test \glspl{ips}. The flexibility of this framework allows it to be used for all kinds of applications where components, such as the positioning engine, can be easily replaced or modified. Beside shoe-box models, other custom rooms can be created and simulated taking into account the materials, reverberation properties, directivities, sample frequencies, air absorption, temperature and SNR and/or SIR values. It was shown that the simulator can mimic a real-life environment well, making it a reliable data source. Future research on acoustic/ultrasonic indoor positioning and classification can make good use of this simulator to train and test \gls{ml} models in addition to traditional algorithms.

\section*{Acknowledgment}
This work was supported by the Research Foundation-Flanders (FWO) under Grant G0D3819N.

{\footnotesize \printbibliography}

\end{document}